\documentclass[11pt,a4paper,superscriptaddress,groupedaddress,nofootnoteinbib]{article} 
\usepackage{jheppub}
\def\({\left(}
\def\){\right)}
\def\[{\left[}
\def\]{\right]}
\def\bea{\begin{eqnarray}}
\def\eea{\end{eqnarray}}
\def\be{\begin{equation}}
\def\ee{\end{equation}}
\def\d{\mathrm{d}}
\def\veck{\mathbf{k}}
\def\vecx{\mathbf{x}}

\makeatletter
\gdef\@fpheader{~}
\gdef\@journal{}
\makeatother
\begin{document}

\title{Gravitational production of scalar dark matter}

\author[a]{Jose A. R. Cembranos,} 
\author[a,b]{Luis J. Garay,}
\author[a]{and Jose M. S\'anchez Vel\'azquez}

\affiliation[a]{Departamento de F\'isica Te\'orica \& IPARCOS, Universidad Complutense de Madrid,\\ 28040 Madrid, Spain}
\affiliation[b]{Instituto de Estructura de la Materia (IEM-CSIC),\\ Serrano 121, 28006 Madrid, Spain}

\emailAdd{cembra@ucm.es}
\emailAdd{luisj.garay@ucm.es}
\emailAdd{jmsvelazquez@ucm.es}

\abstract{We investigate the gravitational production of  scalar dark matter particles during the inflationary and reheating epochs. The oscillatory behavior of the curvature scalar $R$ during the reheating phase generates two different enhancement mechanisms in the particle production. On the one hand, as it has been already discussed in previous works, it induces tachyonic instabilities in the field which are the dominant enhancement mechanism for light masses. On the other hand, we have found that it also provokes a resonant effect in the ultraviolet region of the spectrum which becomes dominant for masses in the range $10^9\,{\rm GeV}$ to $10^{13}\,{\rm GeV}$. We have developed an analytical approximation to describe this resonance effect and its consequences on the ultraviolet regime. Once we have calculated the theoretical gravitational production, we constrain the possible values of the phenomenological field parameters to be considered as a dark matter candidate. We do so by comparing the theoretically predicted abundance with the observed one and ensuring that the theoretical prediction does not lead to overproduction. In particular, we find that there is a region of intermediate masses that is forbidden as they would lead to overproduction.}

\arxivnumber{1910.13937}

\maketitle
\flushbottom




\section{\label{sec:introduction}Introduction}

The nature of dark matter has been an open question since F. Zwicky first proposed its existence to explain the dynamics of the Coma cluster galaxies \cite{DM}. In the last decades, dark matter has become a key ingredient to explain cosmological observations. However, we are still lacking a fundamental description of its nature. The community has followed several approaches to understand it with special effort in looking for extensions of the Standard Model (SM) of particles as many beyond SM proposals include new different fields that could be potential candidates to explain dark matter \cite{Bertone:2004pz}. However, without any conclusive experiment so far, there is great uncertainty in the intrinsic properties of the dark matter candidates as each theoretical proposal predicts different values for its parameters. The only experimental certainty we have so far is that the cross-section of dark matter with the SM fields must be very small \cite{LHCDM,Gaskins:2016cha}. Hence, the question of how it was produced arises naturally. Within this work we study the production of particles due to gravitational effects during both the inflationary and post-inflationary epochs. We focus on discerning if this mechanism can provide enough dark matter density to explain the observed abundance. In particular, we study the case of a scalar field with a non-minimal coupling to the Ricci curvature scalar.

Many different authors have studied the particle production due to gravitational effects in detail. The first works in this arena focused on the production of elementary particles due to the expansion of the Universe \cite{PhysRev.183.1057,birrell1984quantum}. The classic article \cite{PhysRevD.35.2955} analysed the gravitational production of particles after the phase transition to a radiation dominated cosmology after inflation, obtaining a general result for the number density of created particles which is of the order of the energy scale of inflation cube ($n\sim H_0^3$) if the particles can be considered effectively massless during inflation, i.e., if the mass is much smaller than the energy scale during this phase ($m \ll H_0$). More recent papers \cite{Chung:1998zb, Chung:2001cb,Hashiba_2019} have considered the gravitational production as a possible mechanism to produce enough supermassive dark matter particles (WIMPZillas), obtaining then a lower bound for the possible dark matter mass in this scenario. All these works have neglected the importance of the oscillations of the background quantities due to the inflaton dynamics and their impact on the evolution of the quantum field. However, the effects of these oscillatory behaviors on the gravitational production can be important. In \cite{PhysRevD.94.063517,Ema2018}, they analyse the impact of the scale factor oscillations on the gravitational production, and the works \cite{Bassett:1997az, Markkanen:2015xuw} study the tachyonic instability induced on the field by the oscillations of the scalar curvature. There are some recent works which deals with the gravitational production of self-interacting dark matter during the early Universe, imposing some tight constraints on the possible values of the self-coupling parameter \cite{Markkanen_2018, Fairbairn_2019}.

In this work we have focused on a scalar field with negligible interactions with the SM but with a direct coupling to the scalar curvature through a term in the action $\xi R \varphi^{2}$. We start with an initial de Sitter phase which mimics the inflationary dynamics and where the initial vacuum state for the dark matter field is well defined for values of the coupling to the scalar curvature $\xi\geq 1/6$. For couplings $\xi<1/6$ it is known that in de Sitter geometry the vacuum state is unstable \cite{birrell1984quantum}. On the other hand, Ref. \cite{Markkanen:2017adg}  suggests that $\xi$ should not exceed the value  10. Furthermore, analyses based on quantum cosmology  considerations suggest that it should be of order 1 (between 1/6 and 1/3, to be more precise) \cite{Wang:2019spw}. For these reasons together with the numerical difficulties encountered when computing the production for higher values, we will restrict the range of the parameter $\xi$ to lie between the values 1/6 and~1.  Then we have taken into account the oscillatory behavior of the inflaton during the reheating phase and its implications for the curvature scalar. We have seen that besides the already studied tachyonic instability, there is a resonance between the frequency of the field and the frequency of the curvature scalar oscillations that significantly enhances the production of ultraviolet modes for the scalar field. We study this phenomenon using an analytical approximation which allows us to describe only the ultraviolet modes. Comparing the obtained produced gravitational abundance with the observed dark matter abundance, we have set constraints on the possible parameters of the dark matter field in order not to be overproduced. It is important to note that dark matter produced by this mechanism induces adiabatic perturbations instead of isocurvature perturbations as is discussed in \cite{Markkanen:2015xuw,Tenkanen_2019}. Therefore the stringent constraints coming from the CMB observations on the amplitude of isocurvature perturbations \cite{2018arXiv180706211P} do not apply to our study.
 
This manuscript is organized as follows: in section \ref{sec:model} we introduce the model we have considered throughout this work. It consists on a massive scalar field non-minimally coupled to gravity through the Ricci scalar. We are neglecting the back-reaction of the field on the background as its energy density will be small as compared to the one of SM degrees of freedom. Hence, dark matter is considered as a test field on a given background spacetime described by the Friedman-Lema\^itre-Robertson-Walker metric. We are interested in the cosmological production during the early Universe, specifically during the inflationary and reheating epochs. In order to have a well-defined vacuum, we have mimicked the inflationary epoch with the de Sitter solution. Then we consider the reheating dynamics from the oscillations of a massive inflaton which leads to a matter dominated cosmology. We also define what we understand by gravitational production of particles. This production mechanism is based on the time evolution of the vacuum state and the fact that the vacuum defined by an observer at each instant of time does not necessarily match with the evolved vacuum. However, these two states are related by a Bogolyubov transformation and the coefficients of this transformation are related to the produced particle density. In general the mode equation can not be solved analytically so, in section \ref{sec:numerical} we have performed a numerical study of a representative section of the field space of parameters, computing the spectral production and the particle density, which will be used later to set constraints on the dark matter parameters based on the predicted abundance. In section \ref{sec:analytical} an analytical approach is followed to understand the resonance mechanism for the ultraviolet modes that we have observed in the numerical results. Finally, in section \ref{sec:constraints} we show and discuss the numerical results for the production spectra of particles at the end of reheating, and the dark matter abundance that is observed nowadays. We discuss the constraints on the parameter space of the scalar field in order to have a viable dark matter candidate.




\section{\label{sec:model}Background, field dynamics and gravitational production}

We consider a massive scalar field $\varphi$ as a dark matter candidate. Its dynamics in a given curved spacetime is encoded in the action
\be\label{eq:action}
    \mathcal{S}=-\frac{1}{2}\int \d^4x\sqrt{-g}\left(\partial_\mu\varphi\partial^\mu\varphi+m^2\varphi^2+\xi R \varphi^2\right),
\ee
where $g$ is the determinant of the spacetime metric (with signature $(-+++)$), $R$ is the Ricci scalar curvature, $\xi$ is the non-minimal dimensionless coupling constant to gravity, and $m$ is the mass of the dark matter field $\varphi$. In this action we have not included any direct coupling of the dark matter field to the SM sector due to the experimental constraints on the corresponding cross-sections. However, as long as these interactions exist, no matter how weak they are, the renormalization group flow will generate a non-minimal coupling to spacetime curvature \cite{Herranen:2015ima, Markkanen:2018bfx} and therefore we have included it in the action.  Note that this action shares some common terms with the action for Higgs inflationary models due to the presence of the non-minimal coupling $\xi$. However, there is an important difference between them as the Higgs field has quartic self-interactions in its potential while our action is just  quadratic, i.e., it only includes the mass and non-minimal coupling terms. This quartic interaction term is definitively a game changer in the sense that makes the whole dynamics entirely different and hence the role that these two different fields play in the cosmological models.

In cosmology, spacetime geometry is well described in average by the FLRW metric. For later convenience, we write it in conformal time $\eta$. Furthermore, both for simplicity and in accordance with observations \cite{refId0}, we are considering flat spatial sections. Therefore the spacetime geometry will be described by
\be
	\d s^2 = a^2(\eta)(-\d \eta^2 +\delta_{i j} \d x^i\d x^j),
\ee
where the functional form of the scale factor $a(\eta)$ is determined by the energetic content of the Universe during each cosmological epoch. From the point of view of cosmological particle production, the relevant epochs are inflation and reheating. We model the inflationary epoch using the de Sitter solution which is a good first approximation to single-field inflation except for the transition to reheating. On the other hand, during reheating, the behavior of the scale factor is determined by the potential of the inflaton field $\phi$ near its minimum value. In the case at hand the potential under consideration is
\be\label{eq: potential}
V=\frac{1}{2}m_\phi^2\phi^2,
\ee
with $m_\phi$ being the mass of the inflaton field. 
With this potential, it is well known that the background geometry will behave in average as a matter dominated Universe \cite{Kofman:1997yn}. For our study, it suffices to deal only with the average behavior of both the scale factor and the Hubble parameter, for which we keep the symbols $a(\eta)$ and $H(\eta)$. On the other hand, we will take into account the oscillating terms in the curvature scalar sourced by the dynamics of the inflaton. As we will discuss, the gravitational particle production is enhanced by the instabilities provoked by these oscillations in the effective mass term of the field.

If we set the end of inflation at $\eta=0$, the averaged scale factor for these two epochs can be modelled as \cite{PhysRevD.48.647}
\begin{align}\label{eq: scale factor}
a(\eta)&=\begin{cases}
\displaystyle \frac{1}{1-H_{0}\eta}, \quad &\eta\leq0,\\
\displaystyle\(1+\frac{H_{0}\eta}{2}\)^2, \quad &\eta>0,\end{cases}
\end{align}
where $H_{0} = \sqrt{ 12\pi} m_\phi$ is the value of the Hubble parameter at the onset of the single-field inflationary epoch for the model we are considering \eqref{eq: potential}. 

Although this parameterization provides both a continuous scale factor and a continuous Hubble parameter $H(\eta)$ for all values of $\eta$, it fails to provide a continuous Ricci scalar at $\eta=0$. For the approximations we are going to use to solve the mode equations for the field we will need $R$ to be at least a $\mathcal{C}^4$ function. To construct such a function we used an eight order polynomial to interpolate between the value of $R$ at the  end of the de Sitter phase and the one coming from the dynamics of the inflaton field during reheating. We start the interpolation then at $\eta=0$ and finish it at $\eta_{\rm{osc}}$, which is defined as the time in which the Hubble parameter from the background parameterization \eqref{eq: scale factor} coincides with the averaged one from the considered dynamics of the inflaton.

The dynamics of the inflaton is governed, in cosmological time (related to conformal time via $\d t=a(\eta)\d\eta$), by the equation of motion
\be\label{eq: inflaton}
	\ddot{\phi}+3H(t)\dot{\phi}+m_\phi^2\phi=0.
\ee
Inflation ends when $H(t)\sim m_\phi$ and hence during reheating the friction term will be subdominant with respect to the mass term. Henceforth, the solution for the inflationary field can be approximated at first consistent order in a Wentzel-Kramers-Brillouin (WKB) expansion as
\be\label{eq: oscillating inflaton}
    \phi(t) = \frac{\Phi_0}{t}\sin(m_\phi t)\left[1+\mathcal{O}\(\frac{1}{m_\phi t}\)\right],
\ee
where $\Phi_0$ is the amplitude of the field at the end of inflation and depends on the specific inflationary model. In the case of massive chaotic inflation that we are analysing, the observations of the CMB fluctuations \cite{refId0} set the following values for both the mass and the amplitude of the inflaton field at the end of inflation:
\be
m_\phi\simeq1.2\times10^{13} \,{\rm{GeV}}, \qquad	\Phi_0 \simeq \frac{4\sqrt\pi}{\kappa^2},
\ee
where $\kappa^2=8\pi M_P^{-2}$, and $M_P$ is the Planck mass $M_P\simeq 1.22\times 10^{19}\,{\rm{GeV}}$.

\begin{figure}
\centering
\includegraphics[width=.65\columnwidth]{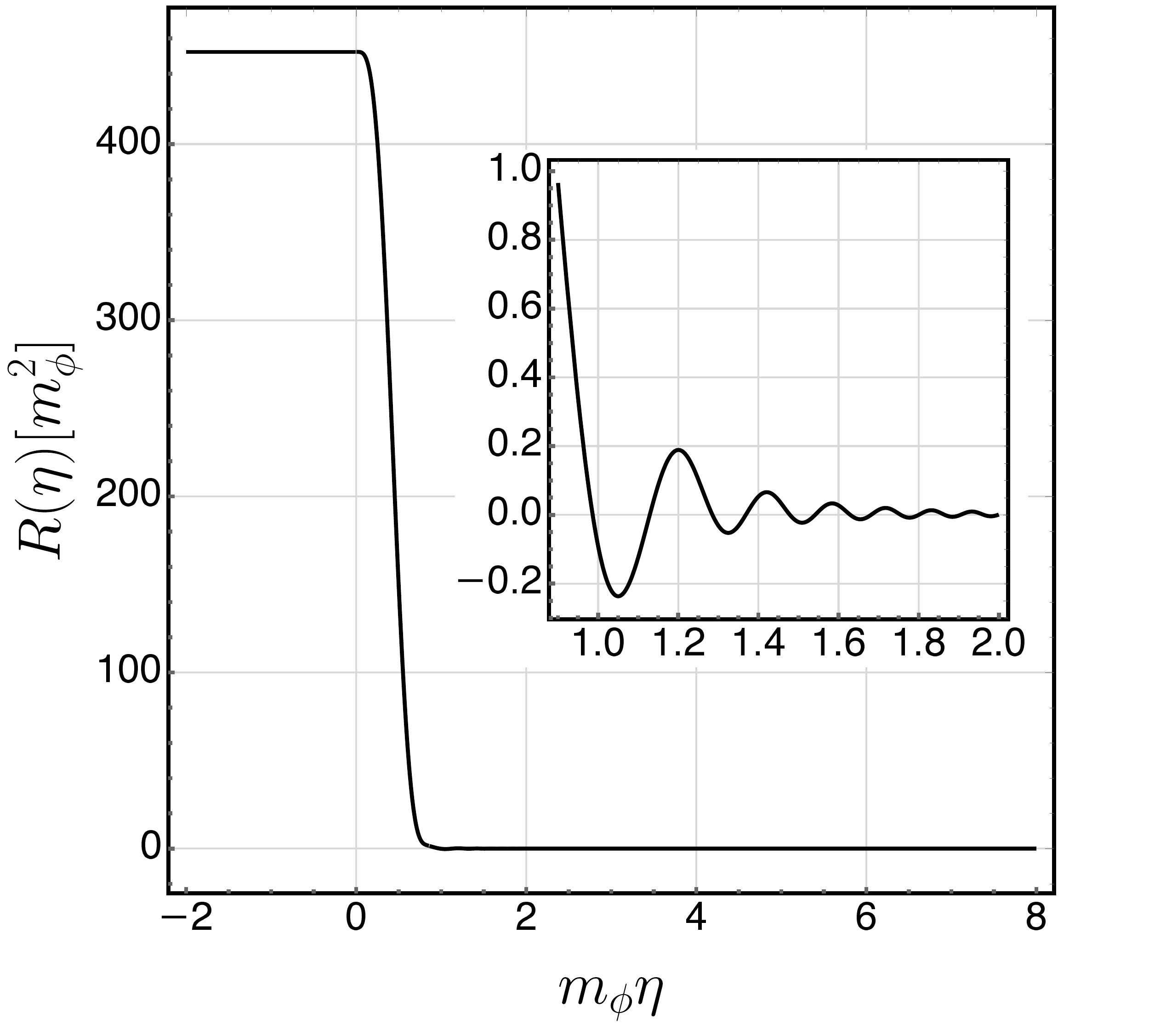}
\caption{Curvature scalar $R(\eta)$ as a function of conformal time. During de Sitter inflation ($\eta<0$) the Ricci scalar remains constant, which is in good agreement with all the inflationary models. The most appreciable difference between our approach and a realistic inflationary model should occur near $\eta=0$. We have modeled the transition between the end of inflation and the onset of oscillations using a polynomial function so the scalar curvature is a $\mathcal{C}^4$ function. For $\eta\geq\eta_{\rm{osc}}$ we have considered the curvature sourced by the oscillations of the inflaton field \eqref{eq: Ricci}.}\label{fig:reta}
\end{figure}

The dynamics of spacetime during the considered epochs is dominated by the energy density of the inflaton field. Hence, the behavior of the curvature scalar  during reheating is given by the trace of the Einstein equations sourced by the energy-momentum tensor of the inflaton:
\be
    R_{\rm{osc}} = \kappa^2(2m_\phi^2\phi^2 -\dot{\phi}^2).
\ee
Using the solution for the inflaton field \eqref{eq: oscillating inflaton}, and keeping all the terms up to the first order WKB approximation, the curvature scalar can be written as
\be\label{eq: Ricci}
 \small  R_{\rm{osc}} \simeq \frac{\kappa^2 m_\phi^2 \Phi_0^2}{t^2}\left[2\sin^2(m_\phi t) -\left(\cos(m_\phi t)-\frac{\sin(m_\phi t)}{m_\phi t}\right)^2\right].
\ee
The resulting curvature scalar from the above approximations is shown in figure \ref{fig:reta}.

In cosmological scenarios, the quantization procedure suffers from an ambiguity which can be traced back to the choice of the canonical pair of variables that we are going to   quantize \cite{Cortez:2015mja}. In our case, there is a preferred choice for the canonical pair of variables if one demand that the quantum theory is invariant under the spatial symmetries of the background and that its quantum dynamics admits an unitary implementation \cite{Gomar:2012xn}. This criterion entails that the field to be quantized must be 
\be\label{eq:rescaledfield}
	\chi = a(\eta)\varphi.
\ee
From the action, we can see that the rescaled field satisfies the equation of motion 
\be
	\partial_\eta^2\chi-\triangle\chi+a^2\left[m^2+\(\xi-\frac{1}{6}\)R\right]\chi=0.
\ee
Exploiting the fact that the spatial sections of the background are flat, we can expand the rescaled field in Fourier modes 
\be\chi = \frac{1}{(2\pi)^{3/2}}\int {\d^3\veck}\ \chi_\veck \ e^{i\veck\cdot\vecx},
\ee
that satisfy the equation
\be\label{eq: mode eq}
	\chi''_\veck+\omega_k^2(\eta)\chi_\veck=0,
\ee
where we have defined the time-dependent frequency as
\be
	\omega_k(\eta)^2 := k^2+a(\eta)^2\left[m^2+\left(\xi-\frac{1}{6}\)R(\eta)\right].
	\label{eq:omega}
\ee
Due to the isotropy of the background, we can always choose bases of solutions  of this equation with elements $v_k$ and $v^*_k$ that only depend on $k=|\veck |$.
The Fourier mode amplitude $\chi_\veck$ is a linear combination of the mode functions $v_k$ and $v^*_k$ with creation and annihilation variables $a_\veck$ and $a^*_{-\veck}$:
\be
	\chi_\veck=\frac{1}{\sqrt{2}}(a_\veck v_k + a^*_{-\veck} v^*_k).
\ee
 
Moreover, to preserve the standard Poisson bracket structure for the field and hence for the creation and annihilation variables, the mode functions must be normalized (we follow the convention of \cite{Mukhanov:2007zz}):
\be\label{eq: normalization}
v_k^*\partial_\eta v_k-v_k\partial_\eta v_k^*=2i.
\ee

The field quantization is carried out by promoting the field variable an its canonically conjugate momentum to operators defined through the creation and annihilation operators $\hat{a}_\veck$ and $\hat{a}^\dag_\veck$ acting on a Fock space. These operators satisfy the following commutation relations:
\be
	[a_\veck,a_{\veck'}]=[a^\dag_\veck,a^\dag_{\veck'}]=0, \qquad [a_\veck,a^\dag_{\veck'}]=\delta(\veck-\veck').
\ee
The Fock space is generated by the action of the creation operators on the vacuum state, defined as the state that is annihilated by all the annihilation operators:
\be
	a_{\veck}\left|0\right>=0,\quad \forall \veck.
\ee

In curved spacetimes the selection of the vacuum state is not unique but depends on the choice of the set of mode functions. The definition of the creation an annihilation operators depends on the choice of basis and hence it introduces an ambiguity in the quantization procedure. In general, different choices will define inequivalent quantizations for the same field. We have followed the criterion presented in \cite{Cortez:2015mja, Gomar:2012xn} to select the field variable to be quantized, requiring the unitary implementation of its quantum dynamics. 
 
In cosmology, the gravitational production occurs because of the time evolution of the background geometry. In these scenarios, the vacuum state for a quantum field is not stationary. Particles are produced because the evolved vacuum state  at a given instant of time is in general  an excited state with respect to the instantaneous vacuum defined at that instant of time.

We can define the vacuum state of the scalar field through the initial conditions given in the de Sitter phase. These initial conditions define a set of modes $v_k$ and $v^*_k$ for the solution of \eqref{eq: mode eq}. The general isotropic solution for the mode equation in the de Sitter phase ($\eta<0$) can be written as:
\be\label{eq: hankel}
	\chi_k(\eta)= \sqrt{-k \tilde{\eta}}\left[c_1 H^{(1)}_{\nu}\(-k\tilde{\eta}\)+c_2 H^{(2)}_{\nu}\(-k\tilde{\eta}\)\right],
\ee
where $\tilde{\eta}=\eta+H_{0}^{-1}$, and $H^{(1)}_{\nu}$, $H^{(2)}_{\nu}$ are the Hankel functions \cite{AbraSteg72} of order 
\be
\nu= \left[1/4-m^2/H_{0}^2-12(\xi-1/6)\right]^{1/2}.
\ee
We define the vacuum state of the field as the one determined by the set of mode functions that in the asymptotic past ($\tilde{\eta}\to-\infty$) behave as the positive-frequency plane waves in Minkowski spacetime. This behavior is obtained if we take $c_2=0$ in \eqref{eq: hankel} and normalize the asymptotic expansion of $v_k$ according to \eqref{eq: normalization}. Then, modes defining the initial vacuum have the form (in the de Sitter phase, i.e. for $\eta\leq 0$)
\be\label{eq:inflation mode}
	v_k(\eta) =e^{-\pi \rm{Im}(\nu)} \sqrt{\frac{-\pi\tilde{\eta}}{2}}H^{(1)}_\nu\(-k\tilde{\eta}\),
\ee
and evolve to the reheating epoch (i.e. for $\eta>0$) according to \eqref{eq: mode eq}.

On the other hand, a comoving observer will have a different instantaneous notion of vacuum. We can use the so-called adiabatic prescription to define this instantaneous vacuum,  commonly used in cosmology. This prescription is appropriate and  well defined if the geometry   evolves slowly enough as compared with the characteristic time scales of the field, i.e., whenever the so-called adiabatic condition is satisfied:
\be\label{eq: adiabaticity}
	\left|\frac{\omega_k'}{\omega_k^2}\right|\ll 1.
\ee
This condition will not be fulfilled in general during the reheating phase for all the modes of the field. However, as we are interested in the production during the whole reheating phase, we only need to have a good prescription for the adiabatic vacuum after this phase has ended. Once the Ricci scalar is no longer oscillating, condition \eqref{eq: adiabaticity} holds and we can define the adiabatic vacuum. The modes $u_k(\eta)$ that define the adiabatic vacuum at the time $\eta_0$ instantaneously behave as the positive-frequency plane wave modes for the vacuum in Minkowski spacetime at $\eta_0$:
\bea\label{eq:vacuum}
	&\displaystyle u_k (\eta_0):=\frac{1}{\sqrt{\omega_k(\eta_0)}},\\
	&\displaystyle u'_k(\eta_0):=-\frac{1}{\sqrt{\omega_k(\eta_0)}}\(i\omega_k(\eta_0)+\frac{1}{2}\frac{\omega'_k(\eta_0)}{\omega_k(\eta_0)}\),
\eea
where $\omega_k(\eta)$ is given in \eqref{eq:omega}.

Once the adiabatic vacuum has been defined for each instant of time, we can compare the mode expansion associated with each Fock space by computing the Bogolyubov coefficients. These coefficients define the linear transformation between two different mode expansions evaluated at the same time. Comparing the evolved modes $v_k $ until the time $\eta$, defined through the initial vacuum state, with the set of modes $u_k$ associated with the adiabatic vacuum at each time $\eta$, we obtain a time-dependent Bogolyubov transformation. For computing the particle production, we are only interested in the coefficient $\beta_k$ relating the negative-frequency modes of one expansion with the positive-frequency modes of the other one. This  coefficient   can be easily calculated and turns out to be \cite{birrell1984quantum}:
\be
	\beta_{k} =  \frac1{2i}(u_k v'_k-v_ku'_k)=\frac{1}{2i\sqrt{\omega_k}}\left[v_{k}'+\(i\omega_k+\frac{1}{2}\frac{\omega'_k}{\omega_k}\)v_{k}\right].
\label{eq: beta}\ee
Through the computation of the expectation value for the number operator defined with the instantaneous vacuum modes in the original vacuum, the gravitational produced density of particles can be obtained directly from the $\beta_k$ Bogolyubov coefficient.
Indeed the total number of particles with momentum $\veck$ created at the time $\eta$ is given by $|\beta_k|^2/[2(2 \pi)^3]$, which integrated over all possible momenta and divided by the spatial volume $a^3$, gives
\be\label{eq: particle density}
	n(t) = \frac{1}{2\pi^2 a^3(t)}\int_0^\infty k^2\left|\beta_k(t)\right|^2\d k,
\ee
where we have already used cosmological time instead of the conformal one.

As stated before, we will focus on the computation of the particle density produced at the end of reheating. Afterwards, we will evolve the computed number density taking into account that the evolution of the Universe is isoentropic after this moment. Therefore, the gravitational production of particles is negligible after reheating has ended. With the evolved number density we can compute the predicted abundance for dark matter from our model as it is explained in section \ref{sec:constraints}.




\section{\label{sec:numerical}Numerical analysis}

In general, the computation of the mode functions during the reheating epoch cannot be performed analytically, although as we discuss in section \ref{sec:analytical}, there are certain regions of the parameter space in which we can carry out an analytical approximation. 

In this section we solve \eqref{eq: mode eq} numerically for a wide range of the scalar field parameters. 
We then introduce the numerical solutions for the mode functions $v_k$   in \eqref{eq: beta}, and via \eqref{eq: particle density} we obtain the total particle production.  We present and discuss the numerical results for the particle production spectra obtained in this way for different parameters of the field, namely for different values of the mass $m$ and the coupling to curvature $\xi$. 

Throughout the numerical exploration of the space of parameters for the field,  we have found that there are four different spectral behaviors as the dynamics of the   curvature scalar produces three different phenomena on the particle production as we will discuss. 

On the one hand, its constant value during the de Sitter phase contributes to the effective mass of the field $m^2+12 H_0^2(\xi-1/6)$. If this effective mass is below the energy scale of inflation $H^{2}_{0}$, the production grows but if it is above it, then the production decreases. 

On the other hand, the oscillations of the curvature scalar during reheating produces two new phenomena. Since the frequency of the field is oscillating, there will be resonant effects whenever the two frequencies (the frequency $\omega_k$ itself and the frequency at which it oscillates) are equal. These resonances will be important mostly in the ultraviolet regions of the spectra. Furthermore, these oscillations make the field suffer from tachyonic instabilities, because the effective mass for the field becomes tachyonic for certain times. This instability affects different $k$-bands as the instability amplitude decreases in each oscillation. Let us now discuss how these effects affect the different regions in parameter space.

\begin{figure}\centering
\includegraphics[width=.65\columnwidth]{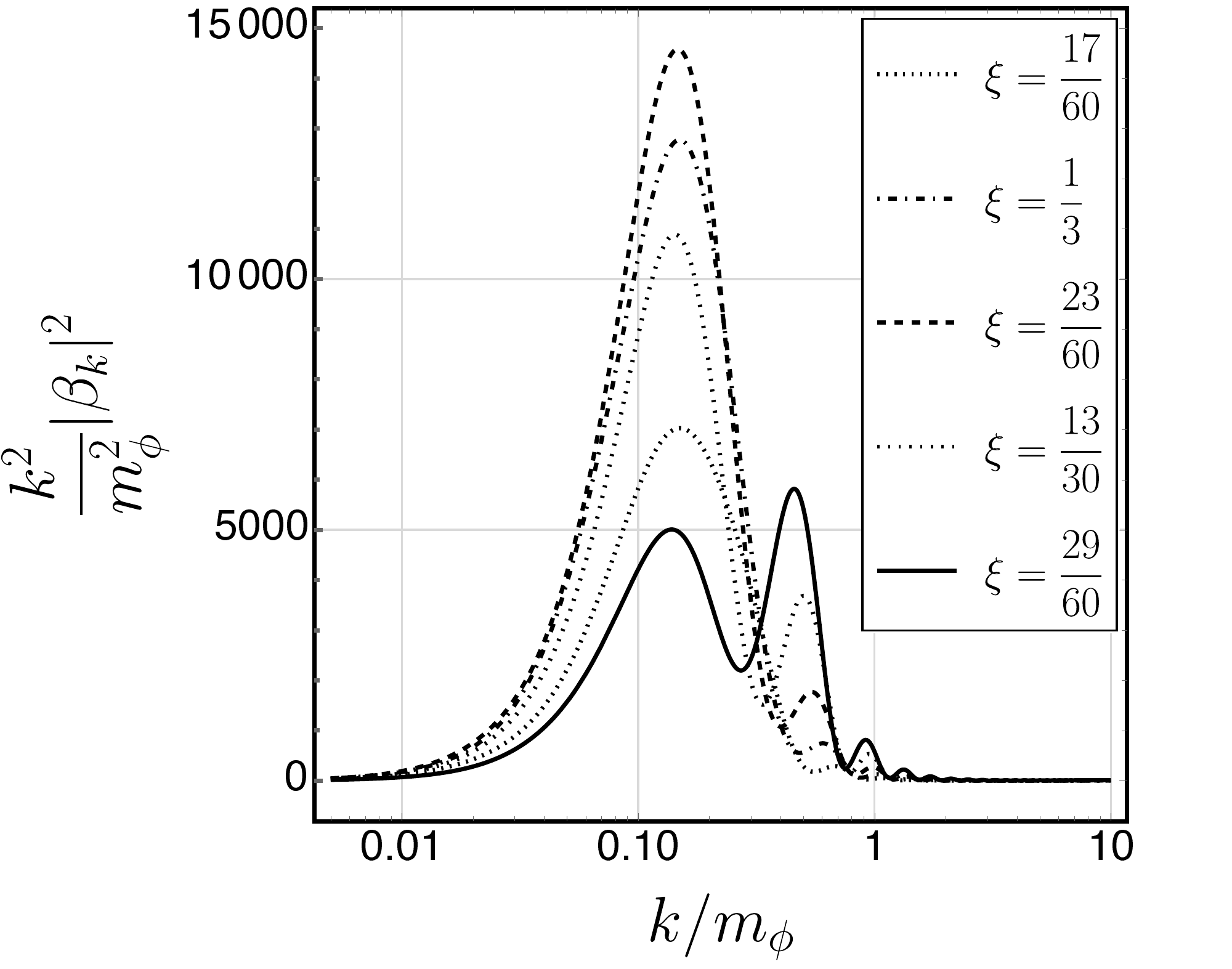}
\caption{Representative spectral particle production in the regime $m^2\ll  (\xi-1/6)|{\rm min}(R_{\rm osc})|$. The specific field mass we are showing is  $m=1.2\times 10^6 \,{\rm{GeV}}$. The different amplified bands appear because each oscillation of the Ricci scalar will excite different modes of the field due to the time dependence of the equations of motion.}\label{fig: spectra light}
\end{figure}

For masses much smaller than the interaction term in the sense that $m^2\ll (\xi-1/6)|{\rm min}(R_{\rm osc})|$, the most important production enhancement mechanism is the tachyonic instability already studied in \cite{Markkanen:2015xuw}. This instability affects mostly the infrared region of the spectrum as it is shown in figure~\ref{fig: spectra light}. As the coupling constant $\xi$ grows, the first infrared peak is damped because the interaction term during the de Sitter epoch makes the effective mass larger and the enhancement occurs on top of the Planckian spectral production at the end of inflation. Furthermore, the larger the interaction term, the larger the ultraviolet peaks, as the instability regions during the oscillatory regime contributes with a larger imaginary effective mass.

\begin{figure}\centering
\includegraphics[width=.65\columnwidth]{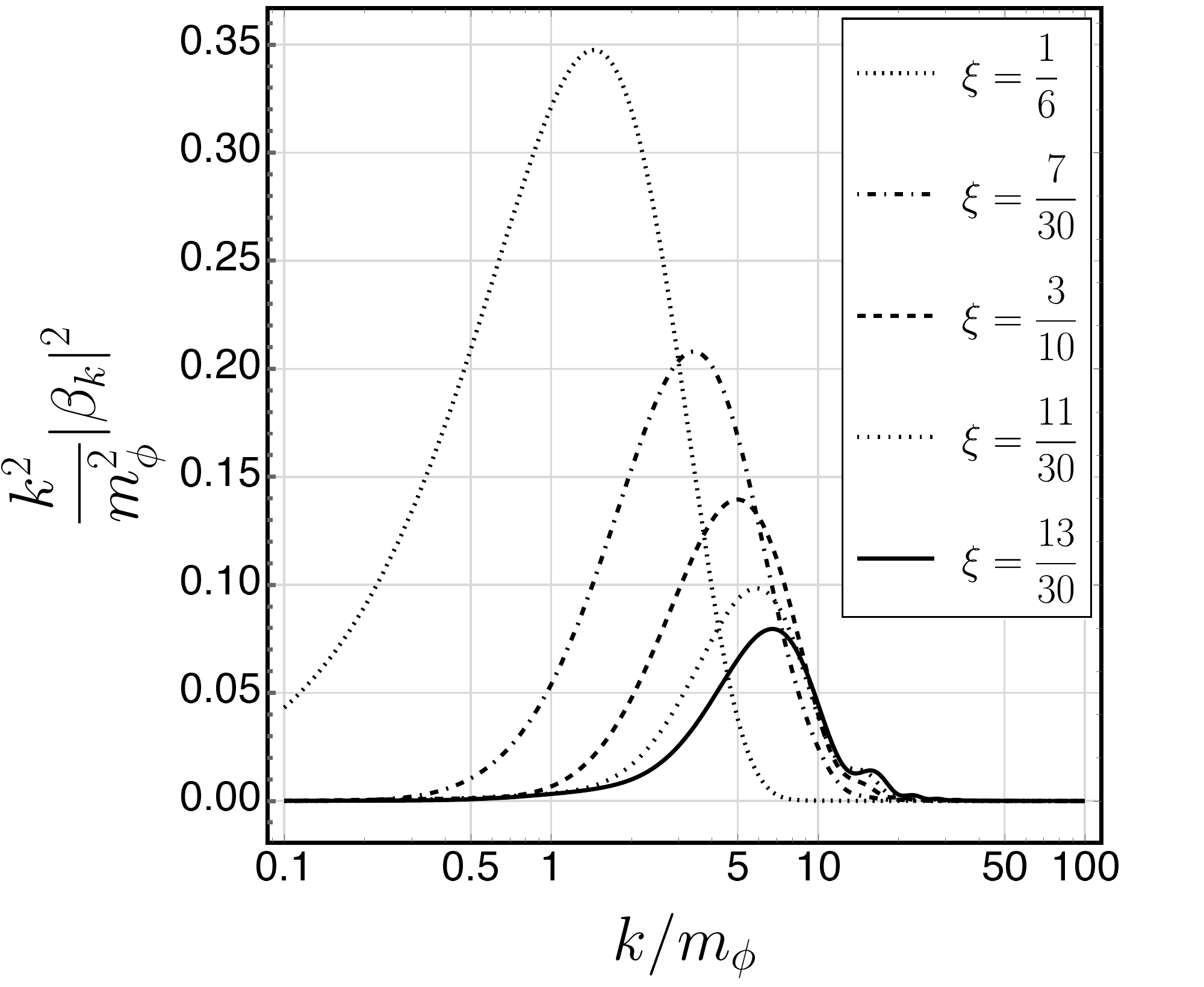}
\caption{Representative spectral particle production for masses in the regime $m>H_{0}$. We are showing the particular field mass $m=1.8\times 10^{13}\,{\rm{GeV}}$. In these cases the production diminishes as the coupling term grows as it contributes to the effective mass during inflation.}\label{fig: spectra large}
\end{figure}

If the mass term is larger than the energy scale of inflation, i.e, $m^{2}>H^{2}_{0}$ and hence larger than the interaction term during the oscillatory phase, then, the resulting spectrum does not differ much from the one obtained in the de Sitter phase. In this scenario, the coupling term increases the effective mass during the inflationary epoch, giving rise to a different Planckian spectrum with no instability-enhancements as it is shown in figure \ref{fig: spectra large}. Therefore, in this regime, as the coupling increases, the particle production decreases.

\begin{figure}\centering
\includegraphics[width=.65\columnwidth]{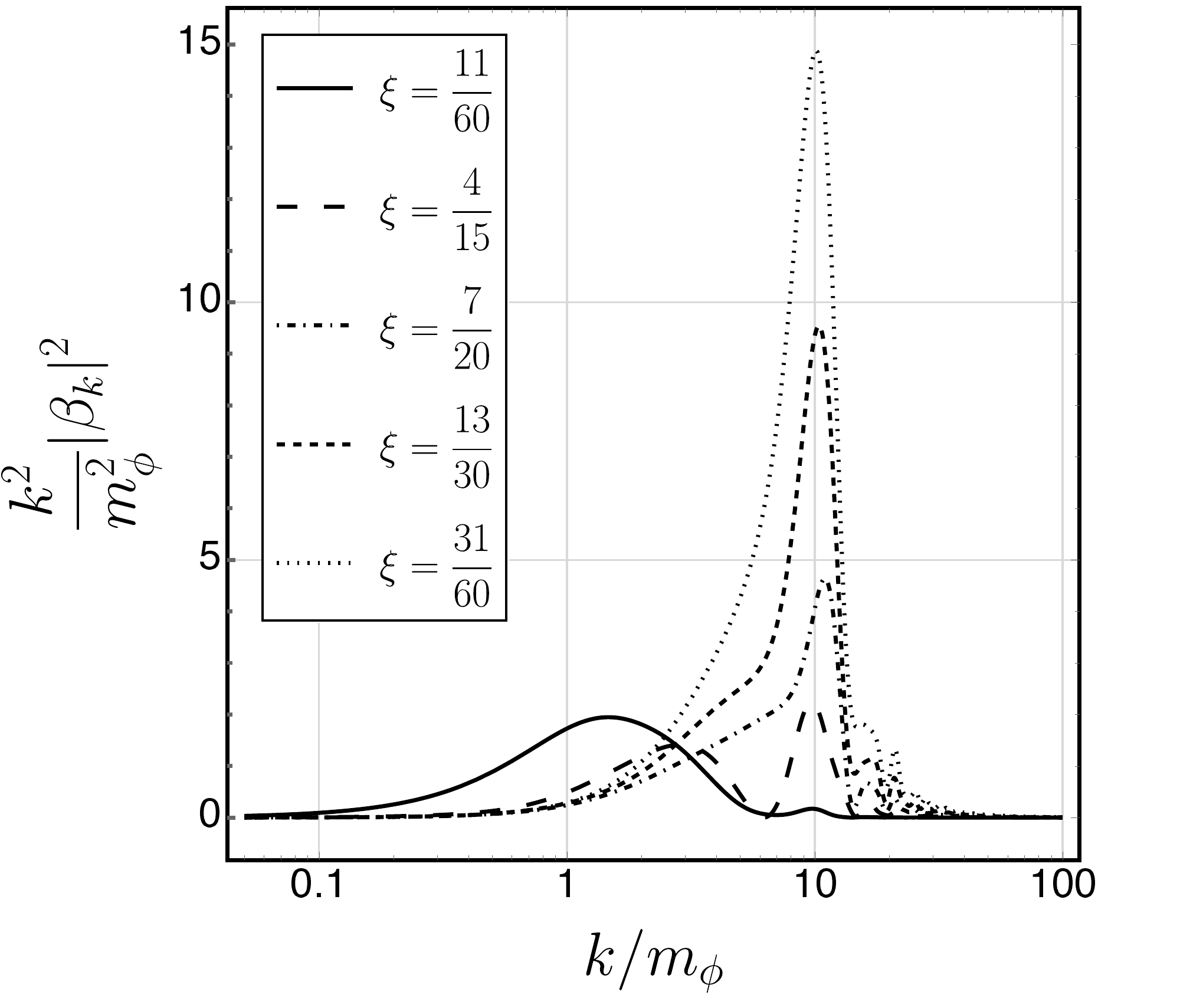}
\caption{Representative spectral particle production for masses in the regime $10^9\,{\rm{GeV}}<m<H_0$. We are showing the particular case for $m=7\times 10^{11}\,{\rm{GeV}}$. In these cases the spectral production increases with the coupling term as the resonance effect becomes more important.}\label{fig: spectra 3rd}
\end{figure}

There is a third regime, shown in figure \ref{fig: spectra 3rd}, in which the mass is below the energy scale of inflation and it is of the order of the coupling term. Within this regime, increasing the coupling constant increases the spectral production as long as the effective mass during inflation is still below the energy scale of inflation. Furthermore, there are resonances due to the fact that the frequency at which the field oscillates coincides at some times with the frequency at which the frequency itself oscillates. This occurs in the ultraviolet region of the spectrum.

\begin{figure}\centering
\includegraphics[width=.65\columnwidth]{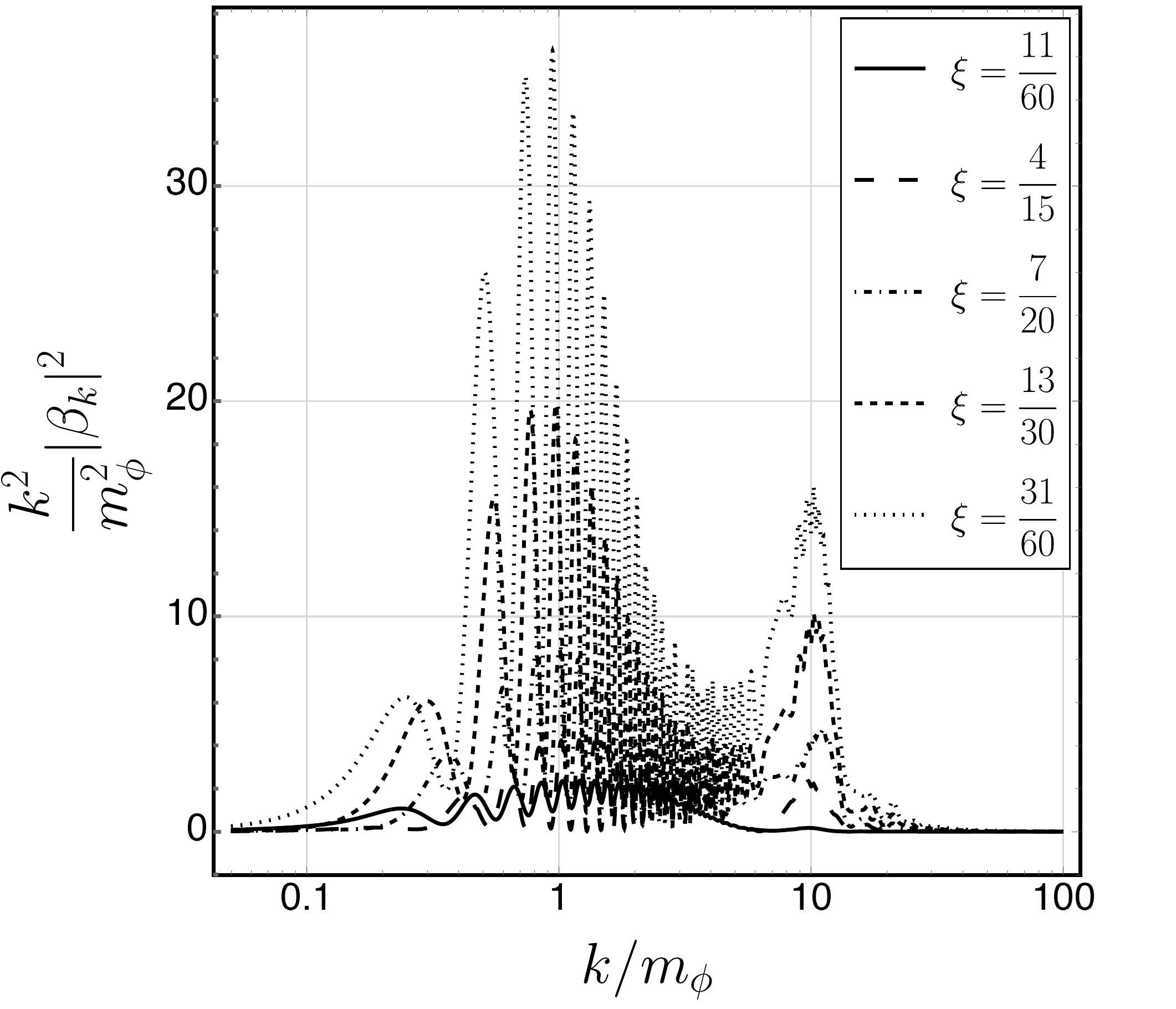}
\caption{Representative spectral particle production for intermediate masses. We are showing the specific field mass $m=3.8\times 10^9 \,{\rm{GeV}}$. In this regime, there are contributions of both the tachyonic instability and the resonant effect with the frequency of the Ricci scalar.}\label{fig: spectra medium}
\end{figure}

Finally, there is a fourth regime for masses $m\sim 10^{9}\,{\rm{GeV}}$ in which there is a transition between the region dominated by the tachyonic instability and the resonance dominated one. The typical spectra for this transition regime is shown in figure \ref{fig: spectra medium}.

\begin{figure}\centering
\includegraphics[width=.65\columnwidth]{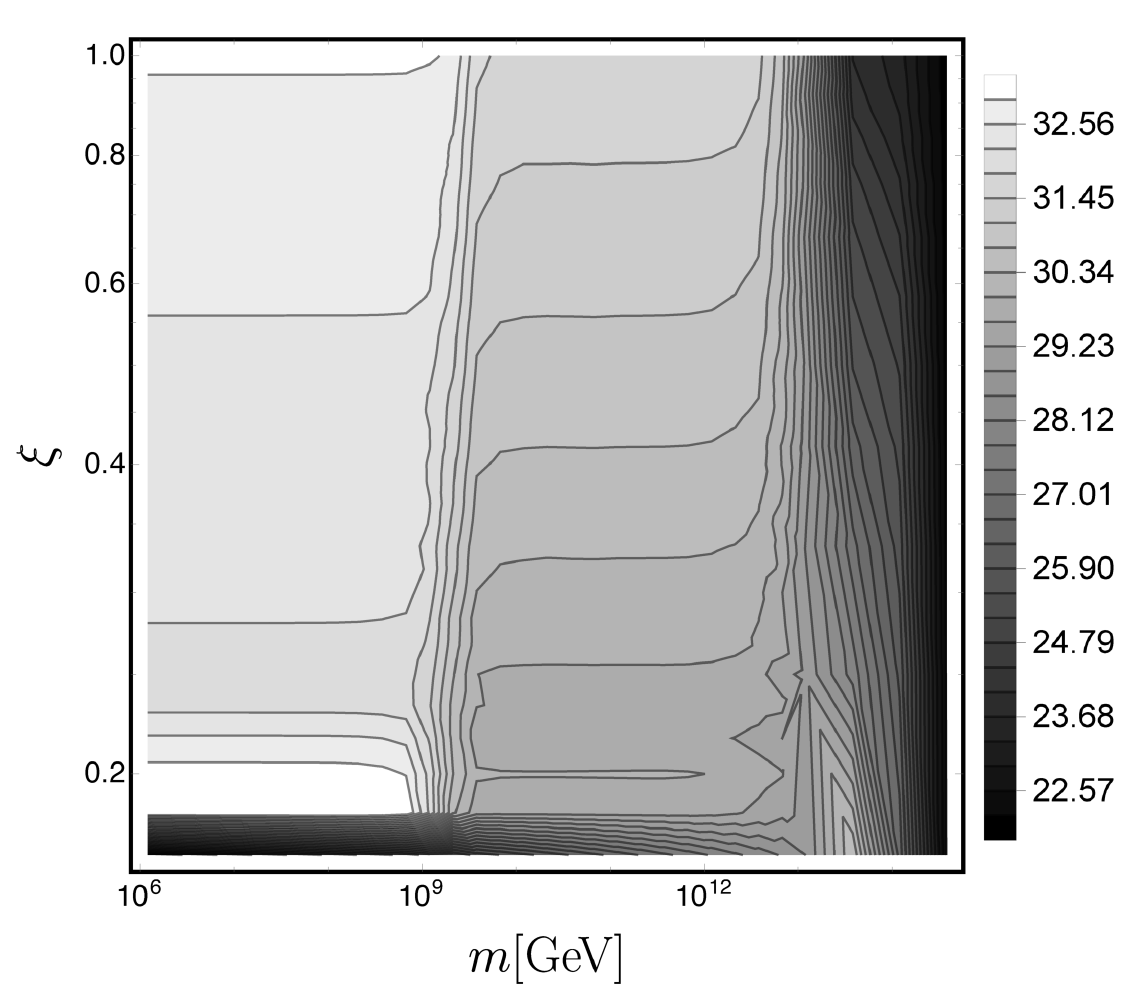}
\caption{In this figure we are showing the particle density in logarithmic scale at the end of reheating, i.e., the contours of $\log_{10} n(t_{\rm{rh}})$ as a function of the mass $m$ and the coupling to curvature $\xi$. We can see clearly the different behaviors of the production. For the lightest masses, the enhancement mechanism is the tachyonic instability. There is a transition regime for masses of about $10^9\,{\rm{GeV}}$ in which there are contributions from the tachyonic instability and the resonant effect. For masses lower than the energy scale of inflation the only enhancement mechanism is the resonance between the frequency of the field and the one of the Ricci scalar. For masses above the energy scale of inflation, there are no enhancement mechanisms and the typical production from a de Sitter phase is recovered.}\label{fig: reheating production}
\end{figure}

Once we have obtained the spectral production density, we can integrate it to obtain the produced particle density at the end of reheating following Equation \eqref{eq: particle density}. In figure \ref{fig: reheating production} this production is shown as a function of the field parameters $m$ and $\xi$. In this plot, we can distinguish the four regions discussed above. The first one is situated in the lightest masses, where the particle production depends only on the value of the coupling to the scalar curvature except for coupling constants in the neighbourhood or the conformal one. The dominant production mechanism in this region is the tachyonic instability induced by the oscillations of the scalar curvature. Around $m\sim 10^9\,{\rm{GeV}}$ there is a transition to the resonant enhancement dominated region. In the range from $m\sim 10^{10}\,{\rm{GeV}}$ to $10^{13}\,{\rm{GeV}}$ there is a dependence of the production on both field parameters, with a monotonic growth in $\xi$ and a monotonic decrease with $m$. Finally, once the mass is above the energy scale of inflation and the coupling to gravity is not too large, we recover the usual behaviour for the production, as it decays rapidly with both the mass and the coupling term for the considered range. It is worth noting that for larger couplings in this regime a resonance enhanced region would be reached eventually and the production would raise again monotonically with the coupling constant value.




\section{\label{sec:analytical}Analytical approximations}

The differential equation for the modes \eqref{eq: mode eq} can be approximated in two different scenarios: either for masses well above the energy scale of inflation $m\gg H_{0}$ or in the ultraviolet regime of the spectrum for all the parameter space.

\subsection{Large mass}

In the case in which the mass of the field is well above the energy scale of inflation ($m\gg H_0$), we can approximate the solution using a second order WKB expansion. Within this approximation, we assume an ansatz for the mode equation of the form
\be\label{eq: WKB}
	v_k = \frac{1}{\sqrt{W_k}}\(A_k e^{-i\int_0^\eta W_k\d s}+B_k e^{i\int_0^\eta W_k\d s}\).
\ee
Here, $W_k$ is determined by introducing this ansatz in the equation of motion for the modes \eqref{eq: mode eq}. If we can neglect the fourth and higher order derivatives of the field frequency, i.e., we can keep the approximation up to second order, this function turns out to be
\be
	W_k^2 = \omega_k^2-\frac{1}{2}\[\frac{\omega''_k}{\omega_k}-\frac{3}{2}\(\frac{\omega'_k}{\omega_k}\)^2\].
\ee
On the other hand, $A_k$ and $B_k$ are coefficients obtained demanding that the modes are $\mathcal{C}^1$ functions in the transition from the de Sitter phase to the reheating one:
\bea\label{eq: coefficients}
	A_k & = & \frac{i}{2\sqrt{W_k}}\[v'^{\rm{dS}}_k -\(i W_k - \frac{W'_k}{2W_k}\)v^{\rm{dS}}_k \] \Bigg|_{\eta=0},\\
	B_k & = & -\frac{i}{2\sqrt{W_k}}\[v'^{\rm{dS}}_k +\(i W_k + \frac{W'_k}{2W_k}\)v^{\rm{dS}}_k \] \Bigg|_{\eta=0},\;
\eea
with $v^{\rm{dS}}_k$ being the mode solution during the de Sitter era, given by Eq. \eqref{eq:inflation mode}.\\

The Bogolyubov coefficient encoding the spectral production can be straightforwardly computed by introducing the WKB approximation for the modes in \eqref{eq: beta}, and it turns to be 
\bea\label{eq: beta WKB}
	&\displaystyle \beta^{\rm{WKB}}_k = -\frac{e^{- i \int_{0}^{\eta} W_{k} \d s}}{4 W_{k}^{3/2} \omega_{k}^{3/2}}\Big\{2W_{k}^{2}\omega_{k}\(A_{k}-B_{k}e^{2i \int_{0}^{\eta} W_{k} \d s}\) \nonumber\\
	 &-\[2W_{k}\omega_{k}^{2}+i(\omega_{k}W'_{k}-\omega'_{k}W_{k})\] \(A_{k}+B_{k}e^{2i \int_{0}^{\eta} W_{k} \d s}\) \Big\}.
\eea

We should note that had we considered the lowest order of the WKB expansion, i.e., $W_k=\omega_k$, the expression \eqref{eq: beta WKB} would has been trivial. Indeed, the mode expansion describing the field in this case would be the same one which defines the adiabatic vacuum at each instant of time and hence the only spectral production would be the one due to the change of vacuum at the end of inflation, encoded in the $B_k$ coefficients of the mode expansion \eqref{eq: WKB}. Hence, the effects of the evolution of the background during reheating would be neglected.

We can obtain an analytic approximation to the number density of particles in this regime by integrating the spectral particle production for all the modes, and dividing by the spatial volume (see \eqref{eq: particle density}):
\begin{align}
 n^{\rm{WKB}} &=\displaystyle\frac{1}{2 \pi^2 a^3(t)}\int_0^\infty \d k\, k^2 \Bigg|\frac{1}{4 W_{k}^{3/2} \omega_{k}^{3/2}}
\bigg\{2W_{k}^{2}\omega_{k}\(A_{k}-B_{k}e^{2i \int_{0}^{\eta} W_{k} \d s}\) \nonumber\\
	 &-\[2W_{k}\omega_{k}^{2}+i(\omega_{k}W'_{k}-\omega'_{k}W_{k})\]
	  \(A_{k}+B_{k}e^{2i \int_{0}^{\eta} W_{k} \d s}\) \bigg\}
\Bigg|^2.
\end{align}

In this expression it is simple to see that in order to include the non-trivial effects of the background evolution in the number density, we need to approximate the solution to \eqref{eq: WKB} with a higher order WKB approximation that the one used to define the adiabatic vacuum (otherwise $W_k=\omega_k$). The integration in modes can be performed with any numerical method and the agreement with number density obtained from the numerical solutions to \eqref{eq: WKB} is exceptionally good.

\subsection{Ultraviolet regime}

\begin{figure}\centering
\includegraphics[width=.65\columnwidth]{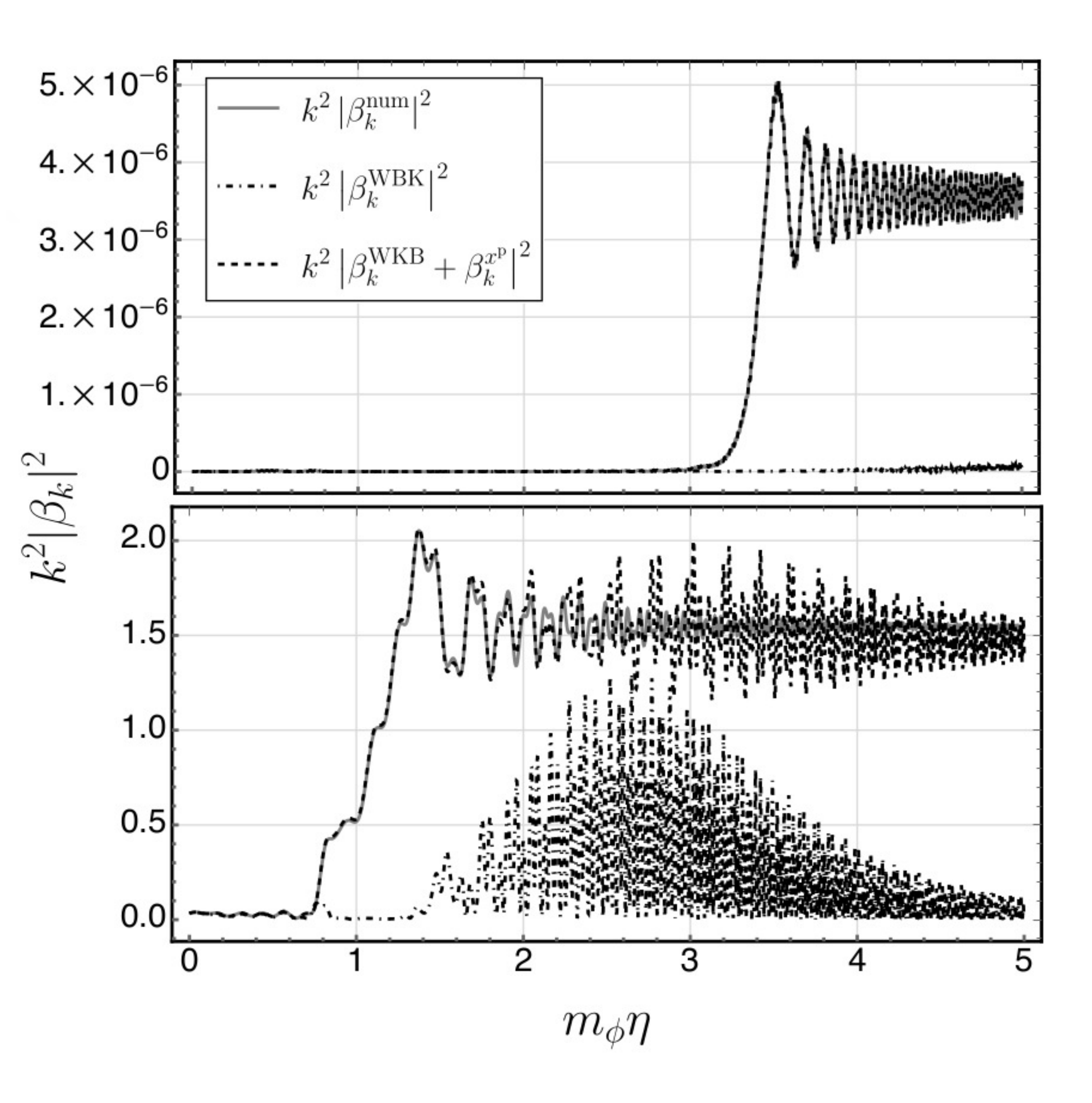}
\caption{Particle production as a function of time for different kinds of computation: numerical, WKB, and refined WKB. In this figure, $\xi=1/4$ and $m=0.1 m_\phi$. The upper panel is calculated for $k=100 m_\phi$ and the lower panel for
 $k=10 m_{\phi}$. 
The WKB approximation (black dot dashed line at the bottom) does not capture the resonance present in the numerical result (gray continuous line) while the refined WKB approximation (black dashed line) agrees with the numerical results.}\label{fig: betaWKB}
\end{figure}

Let us now study the ultraviolet regime ($k\gg m_\phi$). In this case the numerical calculation shows a significant increase in the particle production at the time when the time dependent frequency $\omega_k$ becomes comparable with the characteristic frequency at which it oscillates. The bare WKB approximation does not capture this resonance as shown in figure \ref{fig: betaWKB}. 

In order to analytically deal with this resonance, we have followed a different approximation scheme for this case.
Let us start by defining an auxiliary set of mode functions $x_k$ defined as the difference between the exact solution and the WKB approximation:
\be
	x_k = v^{{\rm{exact}}}_k-v_k^{{\rm{WKB}}}.
\ee
These modes are also normalized according to \eqref{eq: normalization} and satisfy a forced time-dependent harmonic oscillator equation
\be\label{eq: x}
	x''_k+\omega_k^2 x_k = F_k,
\ee
where the source term turns out to be:
\be
	F_k = \frac{v_k^{{\rm{WKB}}}}{4 W_k^{5/2}}\(-3W_k'^2+4W_k^4-4W_k^2\omega_k^2+2W_k W''_k\).
\ee
The reason to define this new set of modes is to extract the resonant behavior explicitly in the equations of motion for the modes and convert it into a classical parametric resonance induced by the source term. 
This source term has two different frequencies: the one from the mode function oscillatory behavior and the one of the oscillation frequency of the scalar curvature. 
The general solution to the equation \eqref{eq: x} can be written as
\be
	x_k=b_1 x^{\rm{h}}_k + b_2 x_k^{{\rm{h}}*}+x^{\rm{p}}_k,
\ee
where $x^{\rm{h}}_k$ is a solution to the homogeneous equation and $x^{\rm{p}}_k$ is a particular solution of the full equation, $b_1$ and $b_2$ being arbitrary constants.
This particular solution can be written as
\be
	x^{\rm{p}}_k = \frac{1}{2i}\int_0^\eta\[x^{\rm{h}}_k(\tau) x^{\rm{h}*}_k(\eta)-x^{\rm{h}*}_k(\tau) x^{\rm{h}}_k(\eta)\]F_k(\tau)\d\tau.
\label{particularxk}\ee
Before the resonance is triggered, the approximate WKB solution $v_k^{{\rm{WKB}}}$ approximates very well the exact solution $v_k^{{\rm{exact}}}$, i.e. $x_k=0$. Therefore, we set $b_1$ and $b_2$ to zero. 
For the solution $x^{\rm{h}}_k$ to the homogeneous equation, necessary to calculate the particular solution \eqref{particularxk}, 
we can simply use the WKB approximation we were using before because we have now isolated the parametric resonant frequency of interest. 

To summarise, the approximate mode functions can be written, in this refined WKB approximation, in the form
\be
v_k=v_k^{{\rm{WKB}}}+x^{\rm{p}}_k,
\ee
where $x^{\rm{p}}_k$ is given by \eqref{particularxk} and, in this expression,
\be
	x^{\rm{h}}_k = \frac{1}{\sqrt{W_k}}{\rm{exp}}\(-i\int_0^\eta W_k \d s\).
\ee

The Bogolyubov coefficients are linear in the mode functions, so the corrections to the $\beta_k$ coefficients obtained through this refined WKB approximations can be added straightforwardly:
\be\label{eq: resonant approx} 
	\beta_k = \beta_k^{\rm{WKB}}+\frac{1}{2i\sqrt{\omega_k}}\[x'^{\rm{p}}_k-\(i\omega_k-\frac{1}{2}\frac{\omega'_k}{\omega_k}\)x^{\rm{p}}_k\].
\ee

As it is shown in figure \ref{fig: betaWKB} this approximation method gives analytical estimations that agree with the numerical results and capture the resonant enhancement perfectly. Using this analytical approximation we can obtain also the number density of particles by replacing \eqref{eq: resonant approx} into \eqref{eq: particle density}.

\section{\label{sec:constraints}Constraints for Dark Matter}

Once we have computed the gravitational particle production at the end of the reheating phase, we can use this result to set constraints on the field parameters if it was to be a dark matter candidate. The constraints are obtained imposing that the predicted abundance equals the one obtained from the observations. In order to make this comparison, we need to evolve the computed production from the end of reheating to the present day. This evolution is computed taking into account that the scalar field is decoupled from the rest of the constituents of the Universe as we have neglected its interactions. Therefore, the evolution of its particle density is solely due to the isoentropic expansion of the Universe. We can rewrite the time dependence of the number density as a dependence on the temperature of the radiation that fills the Universe. This relation arises from the statistical interpretation of the temperature and the evolution of the particle geodesics in a cosmological background \cite{padmanabhan1993structure}. Hence, the evolved number density can be rewritten in terms of the temperature of the background radiation as
\be
	n(t_0) = \frac{a^3(t_{\rm{rh}})}{a^3(t_0)}n(t_{\rm{rh}}) = \frac{g^0_{*S}}{g^{\rm{rh}}_{*S}}\(\frac{T_0}{T_{\rm{rh}}}\)^3 n(T_{\rm{rh}}),
\ee
where $g^{\rm{rh}}_{*S}$ denotes the effective entropic relativistic degrees of freedom of the Standard Model of particles at the end of reheating, $g^{\rm{0}}_{*S}$ denotes the same quantity today, $T_0$ is the temperature today and $T_{\rm{rh}}$ is the temperature at the end of reheating. The actual temperature at which reheating took place is still unknown and depends heavily on the particular model considered. Hence, our constraints will have an uncertainty coming from the indetermination on the specific reheating temperature.\\

\begin{figure}\centering
\includegraphics[width=.65\columnwidth]{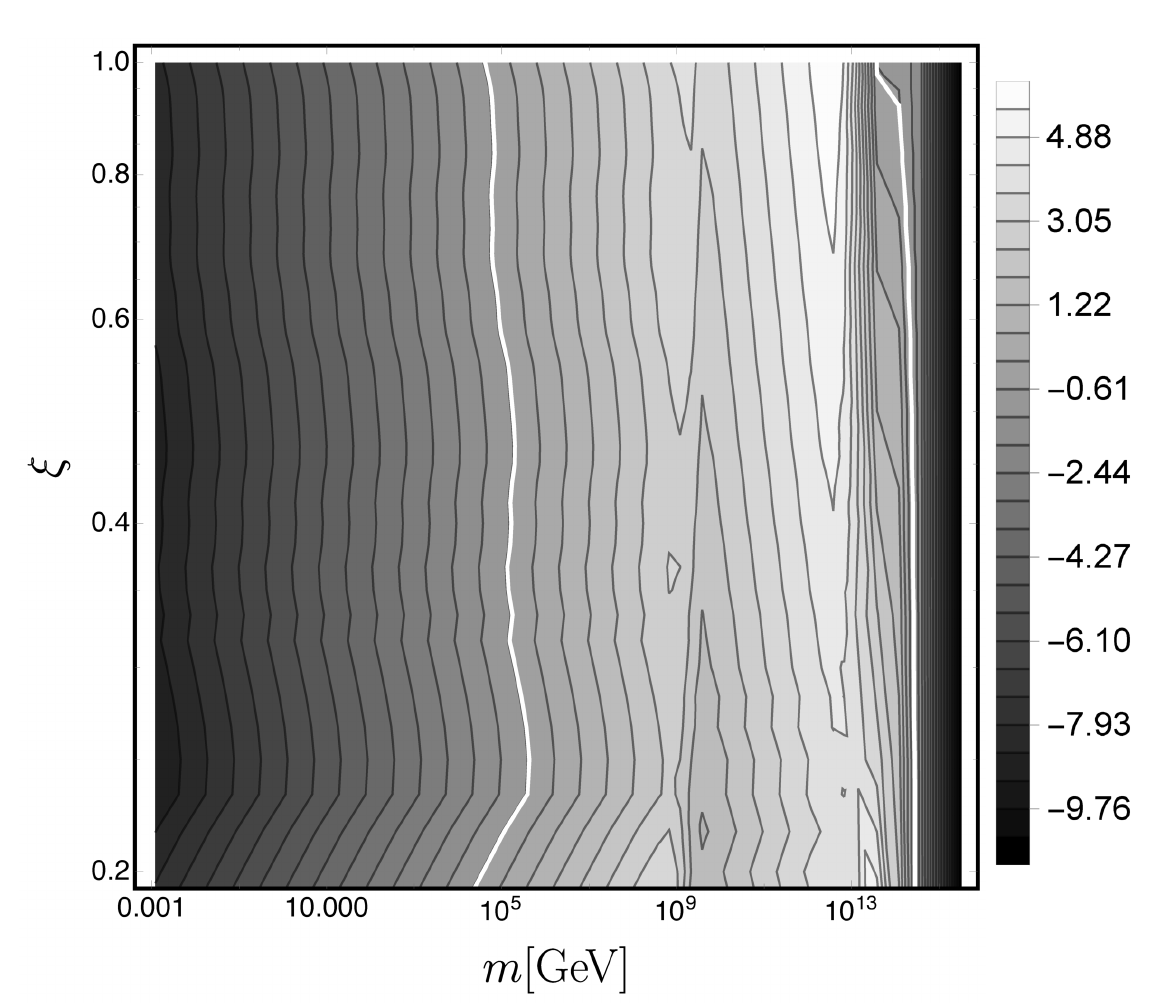}
\caption{We represent the predicted abundance for the scalar field in logarithmic scale, i.e., the contours of $\log_{10}\Omega$ considering the maximum allowed reheating temperature, $T_{{\rm{rh}}} = 10^{15}\,{\rm{GeV}}$. The white contours represents the observed abundance of dark matter $\Omega_{\rm{DM}}=0.268$. The allowed parameter space of the scalar field is to the left of the first white contour and to right of the second one.} \label{fig:phi_abundance}
\end{figure}

\begin{figure}\centering
\includegraphics[width=.6\columnwidth]{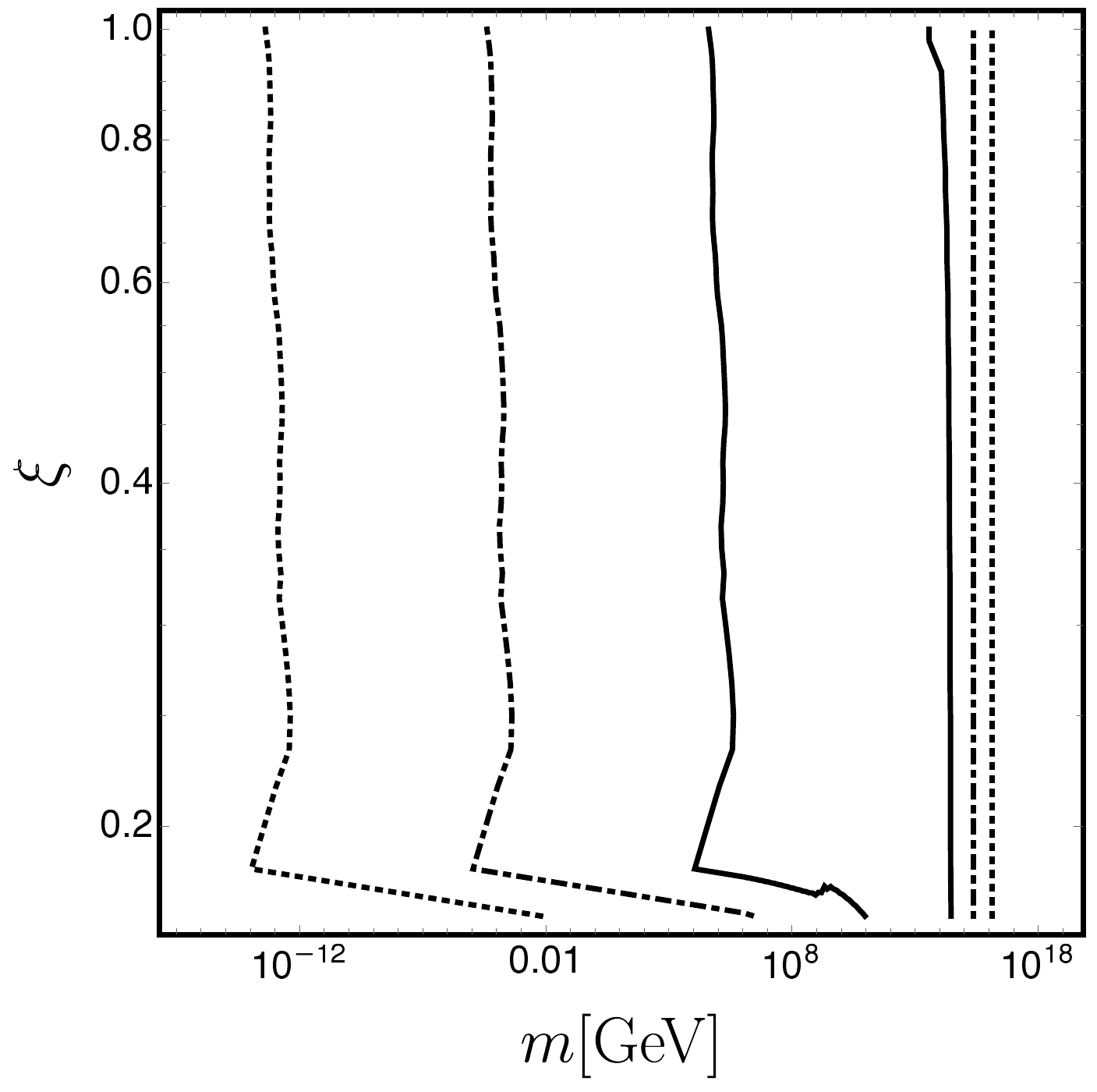}
\caption{The constraints on the field parameter space are represented through the contours corresponding to the observed abundance: $\Omega=0.268$ for different reheating temperatures. The ones obtained for a specific reheating temperature  $T_{\rm{rh}}=10^{15}\,{\rm{GeV}}$ are displayed in solid line. In dot-dashed line we plot the same constraints but for $T_{\rm{rh}}=10^{12}\,{\rm{GeV}}$ and finally they are displayed in dotted line for $T_{\rm{rh}}=10^9\,{\rm{GeV}}$.}\label{fig: allowed}
\end{figure}

Note that we have evolved the number density instead of the energy density of the field. There are two advantages in doing so. On the one hand, we do not need to take into account the effective mass which depends on the curvature term and, on the other hand, we do not need to take into account when each region of the parameter space becomes non-relativistic. Hence, it is simpler to compute its abundance as the energy density today is obtained using just the mass of the particles as they are non-relativistic. The predicted abundance is then obtained through the following expression:
\be
	\Omega(m,\xi) = \frac{\kappa^2}{3H_0^2}\frac{g^0_{*S}}{g^{\rm{rh}}_{*S}}\(\frac{T_0}{T_{\rm{rh}}}\)^3 m\; n(m, \xi),
\ee
where the nought quantities are evaluated at present time and we have emphasized that the density number of particles at the end of reheating depends on the model parameters. The predicted abundance from gravitational production during the early stages of the Universe is shown in figure \ref{fig:phi_abundance} for the maximum allowed value of the reheating temperature: $T_{\rm{rh}}=10^{15}\,{\rm{GeV}}$. This upper bound for the reheating temperature comes from the lack of observation of primordial gravitational waves \cite{PhysRevD.82.023511}.
Although our constraints depend on the reheating temperature, we can set solid constraints if we consider this maximum possible value.
Also in figure \ref{fig:phi_abundance}, we are showing the bounds on the parameter space for the maximum allowed reheating temperature. Note that in this figure we are not showing the region near the conformal coupling to gravity, as the computed abundance depends highly on the coupling value within this region and we have not explored it numerically in sufficient detail. In figure \ref{fig: allowed} we show how the constraints on the parameter space depend heavily on the specific reheating temperature considered. We see that there are two allowed regions to describe dark matter: one for \textit{light} candidates and another one for \textit{heavy} candidates. It is a general result that there is a forbidden mass gap between these two regions for physically meaningful reheating temperatures. This forbidden region is larger as the considered temperature of reheating is lower.

\begin{figure}\centering
\includegraphics[width=.65 \columnwidth]{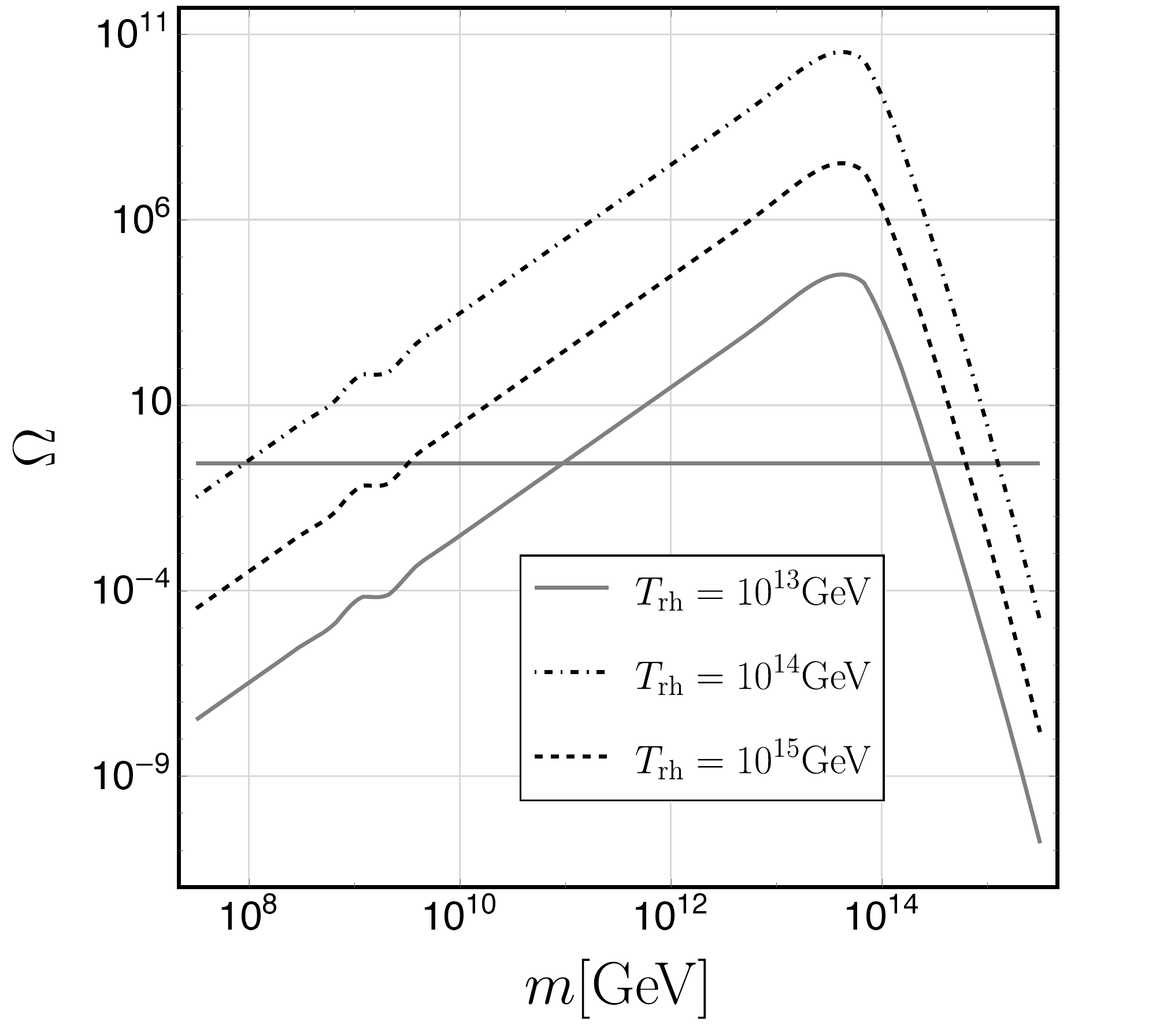}
\caption{Abundance of dark matter for the conformal coupling case ($\xi=1/6$) for different reheating temperatures. The solid horizontal line represents the observed abundance $\Omega=0.268$. The gap between the allowed masses grows larger as the reheating temperature is lower.}\label{fig: conformal}
\end{figure}
We have treated the conformal case in its own and the bounds for the mass in this case are shown in figure \ref{fig: conformal} for different reheating temperatures. It is important to note that, as the production is minimal for the conformal case, the less stringent bounds will come from it. For the maximum allowed reheating temperature, we can see that for the conformal coupling case, the maximum allowed mass for the light sector would be $m\sim 10^{11}\,{\rm{GeV}}$ while for couplings far from the conformal the typical maximum allowed mass would be $m\sim 10^5 \,{\rm{GeV}}$. So in conclusion, the maximum allowed mass for the light candidates sector depends heavily on the coupling constant to curvature $\xi$ due to all the enhancement mechanisms we have discussed throughout this work. Note that the maximum mass coming from the conformal case in this scenario is the upper bound on all the possible masses for the light sector while the minimum mass for this same case is the minimum mass for a WIMPzilla like particle. Furthermore, we can give an empirical law to determine which mass will be the upper bound for the created dark matter particle as a function of the reheating temperature. Fitting the particle production $n$ to an empirical law, we found that it is proportional to the mass in the conformal case. Hence, the upper bound on the mass as a function of the reheating temperature can be written as:
\be
    m^{\rm{light}}_{\rm{max}} \simeq 9.4\times 10^{10}\(\frac{T_{\rm{rh}}}{10^{15}\,{\rm{GeV}}}\)^{3/2}\,{\rm{GeV}}.
\ee

For large enough masses we see that the predicted abundance becomes independent of the coupling and has a maximum value for a mass around the energy scale of inflation. Depending on the temperature at which reheating ended, for high enough masses the abundance can again match the observed one in accordance with \cite{Chung:1998zb}, giving a possible allowed superheavy region. From our computations, this minimum mass would be $m\simeq 5\times 10^{13}\,\rm{GeV}$ for the maximum allowed reheating temperature.

\section{Conclusions}
In this work we have studied the gravitational particle production of a scalar field in the early Universe. We have considered a de Sitter phase prior to the reheating scenario, mimicking the behavior of a geometry sourced by a massive inflationary field. Using this approach we can mimic any inflationary model by adjusting the de Sitter phase parameters and inflaton mass accordingly. We have focused in mimicking a chaotic massive inflation with potential \eqref{eq: potential}. Through this work, we have realized that the oscillations of the curvature scalar $R$ affect the gravitational particle production of a scalar field in different ways. In addition to the already studied tachyonic instability \cite{Bassett:1997az, Markkanen:2015xuw} produced by making the effective mass term imaginary, there is a resonance mechanism that has not been studied before. This resonance affects the ultraviolet region of the spectra and dominates the particle production enhancement for $m>10^9{\rm GeV}$. This resonant behavior is related to the micro-oscillations in the frequency of the field due to the Ricci oscillations, in analogy to what happens in the Mathieu equation \cite{AbraSteg72}. This result is in disagreement with previous works as \cite{Markkanen:2015xuw} where they ignored the effect of having many oscillations of the scalar curvature and they have not taken into account the resonant effect induced by these oscillations.
One of the conclusions of this work is that gravitational production can account for the observed dark matter abundance. Furthermore, this gravitational production has allowed us to set constraints on the possible values of the mass of the dark matter particle if any possible direct coupling with the SM is negligible. In particular we have found that there is a forbidden band of masses which split the dark matter into either light or supermassive candidates. The width of this band is very sensitive to the reheating temperature. We have seen that the largest possible mass for the light candidate is obtained in the conformal coupling case and that for masses above the inflationary energy scale, the production become independent of the coupling constant $\xi$. Hence, analysing the conformal case is sufficient to set the constraints on the possible masses for the scalar dark matter candidate. However, the most stringent constrains come from the cases in which $\xi>1/6$ as the possible maximum mass is much smaller than in the conformal case. For values of the coupling constant $\xi$ not covered in this work, i.e., $\xi>1$, the gravitational production would be further enhanced, widening then the range of forbidden masses for a dark matter candidate. The most affected region would be the large mass one because the field would suffer from tachyonic instabilities for large enough values of $\xi$, enhancing then heavily the production and dragging the allowed lightest heavy mass to larger values. For the small and intermediate mass regions, considering larger values of $\xi$ would increase monotonically the production, dragging the allowed values of the mass for the dark matter candidate to even lighter ones.

\section*{Acknowledgments}
This work has been financially supported in part by the MINECO (Spain) projects FIS2016-78859-P (AEI/FEDER) and FIS2017-86497-C2-2-P (with FEDER contribution). JMSV acknowledges financial support from Universidad Complutense de Madrid through the predoctoral grant CT27/16. This work was made possible by Institut Pascal at Université Paris-Saclay with the support of the P2I and SPU research departments and  the P2IO Laboratory of Excellence (program “Investissements d’avenir” ANR-11-IDEX-0003-01 Paris-Saclay and ANR-10-LABX-0038), as well as the IPhT"

\bibliographystyle{JHEP}
\bibliography{bibliography}

\providecommand{\href}[2]{#2}\begingroup\raggedright\begin{thebibliography}{10}

\bibitem{DM}
F.~{Zwicky}, \emph{{Die Rotverschiebung von extragalaktischen Nebeln}},
  {\emph{Helv. Phys. Acta} {\bfseries 6} (1933) 110}.

\bibitem{Bertone:2004pz}
G.~Bertone, D.~Hooper and J.~Silk, \emph{{Particle dark matter: Evidence,
  candidates and constraints}},
  \href{https://doi.org/10.1016/j.physrep.2004.08.031}{\emph{Phys. Rept.}
  {\bfseries 405} (2005) 279}
  [\href{https://arxiv.org/abs/hep-ph/0404175}{{\ttfamily hep-ph/0404175}}].

\bibitem{LHCDM}
F.~Kahlhoefer, \emph{Review of lhc dark matter searches},
  \href{https://doi.org/10.1142/S0217751X1730006X}{\emph{Int. J. Mod. Phys.}
  {\bfseries 32} (2017) 1730006}
  [\href{https://arxiv.org/abs/1702.02430}{{\ttfamily 1702.02430}}].

\bibitem{Gaskins:2016cha}
J.~M. Gaskins, \emph{{A review of indirect searches for particle dark matter}},
  \href{https://doi.org/10.1080/00107514.2016.1175160}{\emph{Contemp. Phys.}
  {\bfseries 57} (2016) 496}
  [\href{https://arxiv.org/abs/1604.00014}{{\ttfamily 1604.00014}}].

\bibitem{PhysRev.183.1057}
L.~Parker, \emph{Quantized fields and particle creation in expanding universes.
  {I}}, \href{https://doi.org/10.1103/PhysRev.183.1057}{\emph{Phys. Rev.}
  {\bfseries 183} (1969) 1057}.

\bibitem{birrell1984quantum}
N.~Birrell and P.~Davies, \emph{Quantum Fields in Curved Space}, Cambridge
  Monographs on Mathematical Physics. Cambridge University Press, 1984.

\bibitem{PhysRevD.35.2955}
L.~H. Ford, \emph{Gravitational particle creation and inflation},
  \href{https://doi.org/10.1103/PhysRevD.35.2955}{\emph{Phys. Rev.} {\bfseries
  D 35} (1987) 2955}.

\bibitem{Chung:1998zb}
D.~J.~H. Chung, E.~W. Kolb and A.~Riotto, \emph{{Superheavy dark matter}},
  \href{https://doi.org/10.1103/PhysRevD.59.023501}{\emph{Phys. Rev.}
  {\bfseries D 59} (1999) 023501}
  [\href{https://arxiv.org/abs/hep-ph/9802238}{{\ttfamily hep-ph/9802238}}].

\bibitem{Chung:2001cb}
D.~J.~H. Chung, P.~Crotty, E.~W. Kolb and A.~Riotto, \emph{{Gravitational
  production of superheavy dark matter}},
  \href{https://doi.org/10.1103/PhysRevD.64.043503}{\emph{Phys. Rev.}
  {\bfseries D 64} (2001) 043503}
  [\href{https://arxiv.org/abs/hep-ph/0104100}{{\ttfamily hep-ph/0104100}}].

\bibitem{Hashiba_2019}
S.~Hashiba and J.~Yokoyama, \emph{Gravitational particle creation for dark
  matter and reheating},
  \href{https://doi.org/10.1103/PhysRevD.99.043008}{\emph{Phys. Rev.}
  {\bfseries D 99} (2019) 043008}
  [\href{https://arxiv.org/abs/1812.10032}{{\ttfamily 1812.10032}}].

\bibitem{PhysRevD.94.063517}
Y.~Ema, R.~Jinno, K.~Mukaida and K.~Nakayama, \emph{Gravitational particle
  production in oscillating backgrounds and its cosmological implications},
  \href{https://doi.org/10.1103/PhysRevD.94.063517}{\emph{Phys. Rev.}
  {\bfseries D 94} (2016) 063517}
  [\href{https://arxiv.org/abs/1604.08898}{{\ttfamily 1604.08898}}].

\bibitem{Ema2018}
Y.~Ema, K.~Nakayama and Y.~Tang, \emph{Production of purely gravitational dark
  matter}, \href{https://doi.org/10.1007/JHEP09(2018)135}{\emph{J. High Energ.
  Phys.} {\bfseries 2018} (2018) 135}
  [\href{https://arxiv.org/abs/1804.07471}{{\ttfamily 1804.07471}}].

\bibitem{Bassett:1997az}
B.~A. Bassett and S.~Liberati, \emph{{Geometric reheating after inflation}},
  \href{https://doi.org/10.1103/PhysRevD.58.021302}{\emph{Phys. Rev.}
  {\bfseries D 58} (1998) 021302}
  [\href{https://arxiv.org/abs/hep-ph/9709417}{{\ttfamily hep-ph/9709417}}].

\bibitem{Markkanen:2015xuw}
T.~Markkanen and S.~Nurmi, \emph{Dark matter from gravitational particle
  production at reheating},
  \href{https://doi.org/10.1088/1475-7516/2017/02/008}{\emph{J. Cosmol.
  Astropart. Phys.} {\bfseries 2017} (2017) 008}
  [\href{https://arxiv.org/abs/1512.07288}{{\ttfamily 1512.07288}}].

\bibitem{Markkanen_2018}
T.~Markkanen, A.~Rajantie and T.~Tenkanen, \emph{Spectator dark matter},
  \href{https://doi.org/10.1103/physrevd.98.123532}{\emph{Phys. Rev.}
  {\bfseries D 98} (2018) 123532}
  [\href{https://arxiv.org/abs/1811.02586}{{\ttfamily 1811.02586}}].

\bibitem{Fairbairn_2019}
M.~Fairbairn, K.~Kainulainen, T.~Markkanen and S.~Nurmi, \emph{Despicable dark
  relics: generated by gravity with unconstrained masses},
  \href{https://doi.org/10.1088/1475-7516/2019/04/005}{\emph{J. Cosmol.
  Astropart. Phys.} {\bfseries 2019} (2019) 005}
  [\href{https://arxiv.org/abs/1808.08236}{{\ttfamily 1808.08236}}].

\bibitem{Markkanen:2017adg}
T.~Markkanen, \emph{{Vacuum Stability in the Early Universe and the
  Backreaction of Classical Gravity}},
  \href{https://doi.org/10.1098/rsta.2017.0115}{\emph{Phil. Trans. Roy. Soc.
  Lond.} {\bfseries A 376} (2018) 20170115}
  [\href{https://arxiv.org/abs/1707.03415}{{\ttfamily 1707.03415}}].

\bibitem{Wang:2019spw}
S.-J. Wang, M.~Yamada and A.~Vilenkin, \emph{{Constraints on non-minimal
  coupling from quantum cosmology}},
  \href{https://doi.org/10.1088/1475-7516/2019/08/025}{\emph{J. Cosmol.
  Astropart. Phys.} {\bfseries 2019} (2019) 025}
  [\href{https://arxiv.org/abs/1903.11736}{{\ttfamily 1903.11736}}].

\bibitem{Tenkanen_2019}
T.~Tenkanen, \emph{Dark matter from scalar field fluctuations},
  \href{https://doi.org/10.1103/physrevlett.123.061302}{\emph{Phys. Rev. Lett.}
  {\bfseries 123} (2019) 061302}
  [\href{https://arxiv.org/abs/1905.01214}{{\ttfamily 1905.01214}}].

\bibitem{2018arXiv180706211P}
{[Planck Collaboration]}, {Akrami, Y.} et~al., \emph{{Planck 2018 results. X.
  Constraints on inflation}},
  \href{https://arxiv.org/abs/1807.06211}{{\ttfamily 1807.06211}}.

\bibitem{Herranen:2015ima}
M.~Herranen, T.~Markkanen, S.~Nurmi and A.~Rajantie, \emph{{Spacetime curvature
  and Higgs stability after inflation}},
  \href{https://doi.org/10.1103/PhysRevLett.115.241301}{\emph{Phys. Rev. Lett.}
  {\bfseries 115} (2015) 241301}
  [\href{https://arxiv.org/abs/1506.04065}{{\ttfamily 1506.04065}}].

\bibitem{Markkanen:2018bfx}
T.~Markkanen, S.~Nurmi, A.~Rajantie and S.~Stopyra, \emph{{The 1-loop effective
  potential for the Standard Model in curved spacetime}},
  \href{https://doi.org/10.1007/JHEP06(2018)040}{\emph{J. High Energ. Phys.}
  {\bfseries 2018} (2018) 040}
  [\href{https://arxiv.org/abs/1804.02020}{{\ttfamily 1804.02020}}].

\bibitem{refId0}
{[Planck Collaboration]}, {Ade, P. A. R.} et~al., \emph{Planck 2015 results -
  {XIII}. cosmological parameters},
  \href{https://doi.org/10.1051/0004-6361/201525830}{\emph{Astron. Astrophys.}
  {\bfseries 594} (2016) A13}
  [\href{https://arxiv.org/abs/1502.01589}{{\ttfamily 1502.01589}}].

\bibitem{Kofman:1997yn}
L.~Kofman, A.~D. Linde and A.~A. Starobinsky, \emph{{Towards the theory of
  reheating after inflation}},
  \href{https://doi.org/10.1103/PhysRevD.56.3258}{\emph{Phys. Rev.} {\bfseries
  D 56} (1997) 3258} [\href{https://arxiv.org/abs/hep-ph/9704452}{{\ttfamily
  hep-ph/9704452}}].

\bibitem{PhysRevD.48.647}
M.~R. de~Garcia~Maia, \emph{Spectrum and energy density of relic gravitons in
  flat robertson-walker universes},
  \href{https://doi.org/10.1103/PhysRevD.48.647}{\emph{Phys. Rev.} {\bfseries D
  48} (1993) 647}.

\bibitem{Cortez:2015mja}
J.~Cortez, G.~A. Mena~Marug\'an and J.~M. Velhinho, \emph{{Quantum unitary
  dynamics in cosmological spacetimes}},
  \href{https://doi.org/10.1016/j.aop.2015.09.016}{\emph{Annals Phys.}
  {\bfseries 363} (2015) 36}
  [\href{https://arxiv.org/abs/1509.06171}{{\ttfamily 1509.06171}}].

\bibitem{Gomar:2012xn}
L.~Castell{\'{o}}~Gomar, J.~Cortez, D.~Mart{\'{\i}}n-de Blas, G.~A.
  Mena~Marug{\'{a}}n and J.~M. Velhinho, \emph{Uniqueness of the fock
  quantization of scalar fields in spatially flat cosmological spacetimes},
  \href{https://doi.org/10.1088/1475-7516/2012/11/001}{\emph{J. Cosmol.
  Astropart. Phys.} {\bfseries 2012} (2012) 001}
  [\href{https://arxiv.org/abs/1211.5176}{{\ttfamily 1211.5176}}].

\bibitem{Mukhanov:2007zz}
V.~Mukhanov and S.~Winitzki, \emph{{Introduction to quantum effects in
  gravity}}. Cambridge University Press, 2007.

\bibitem{AbraSteg72}
M.~Abramowitz and I.~A. Stegun, eds., \emph{Handbook of Mathematical Functions
  with Formulas, Graphs, and Mathematical Tables}. U.S. Government Printing
  Office, Washington, DC, USA, tenth printing~ed., 1972.

\bibitem{padmanabhan1993structure}
T.~Padmanabhan, \emph{Structure Formation in the Universe}. Cambridge
  University Press, 1993.

\bibitem{PhysRevD.82.023511}
J.~Martin and C.~Ringeval, \emph{First {CMB} constraints on the inflationary
  reheating temperature},
  \href{https://doi.org/10.1103/PhysRevD.82.023511}{\emph{Phys. Rev.}
  {\bfseries D 82} (2010) 023511}
  [\href{https://arxiv.org/abs/1004.5525}{{\ttfamily 1004.5525}}].

\end{thebibliography}\endgroup
\end{document}